# A Framework of Critical Success Factors for Agile Software Development

Ridewaan Hanslo
University of Cape Town
Cape Town
South Africa
ridewaan.hanslo@uct.ac.za

Maureen Tanner
University of Cape Town
Cape Town
South Africa
mc.tanner@uct.ac.za

## Abstract

Despite the popularity of Agile software development, achieving consistent project success remains challenging. This systematic literature review identifies critical success factors (CSFs) in Agile projects by analyzing 53 primary studies. Employing thematic synthesis with content analysis, our analysis yielded 21 CSFs categorized into five themes: organizational, people, technical, process, and project. Team effectiveness and project management emerged as the most frequently cited CSFs, highlighting the importance of people and process factors. These interpreted themes and factors contributed to the development of a theoretical framework to identify how these factors contribute to project success. This study offers valuable insights for researchers and practitioners, guiding future research to validate these findings and test the proposed framework using quantitative methods.

## 1 Introduction

The Agile Manifesto of 2001 presented an alternative to traditional, sequential software development project management [1][2]. Agile methodologies provide an incremental and iterative software delivery, which allows for a more dynamic development approach with fewer costly errors [2][3]. The adoption of Agile methods like Scrum, Extreme Programming, and Kanban has been gradual and has fostered positive sentiments regarding their use [4]. This sentiment is reflected in project outcomes, where Agile projects are three times more likely to succeed than those using traditional methods like Waterfall [4].

### 1.1 Problem Statement

Despite the advantages of Agile, it faces significant challenges. The success rate for Agile projects is reported to be under 50%, resulting in billions of dollars in economic losses due to failed projects [5]. In response, researchers have focused on identifying the Critical Success Factors (CSFs) that influence Agile project outcomes [1][3][6]-[8]. The CSF theory, developed by Rockart [9], provides a framework for identifying the "limited number of areas in which results, if they are satisfactory, will ensure successful competitive performance for the organization". However, research applying this theory to Agile projects has produced inconsistent findings. For instance, a study by Chow and Cao [7] identified six significant CSFs: delivery strategy, Agile software engineering techniques, team capability, project management process, team environment, and customer involvement.

A replication of that study by Stankovic et al. [8] found different CSFs to be significant: the project management process, project definition process, project nature, and project schedule. Additionally, the quality of Systematic Literature Reviews (SLRs) used to discover these factors is inconsistent. Previous SLRs often lack the necessary information to be replicated, failing to be systematic, explicit, and comprehensive [10]-[13].

### 1.2 Research Objective and Question

The primary aim of this research is to identify the critically significant factors that contribute to the success of Agile software development projects, to improve project outcomes. To accomplish this, the first objective is to perform an SLR that utilizes thematic synthesis and content analysis. This process will generate CSFs by identifying recurring themes across existing studies to develop new insights. This leads to the central research question: "What are the key factors contributing to Agile software development project success?"

## 2  Research Method

This study employs an SLR, which is a structured method for identifying, evaluating, and synthesizing primary research studies [14][15]. The process involves identifying sources, selecting relevant studies, extracting data, synthesizing findings into themes, and reporting the results [14].

### 2.1  Search Strategy

Search strings were developed by combining keywords with logical operators like AND and OR [16]. The creation of these strings is left to the researcher's discretion to find the most appropriate list [16][17]. The comprehensive search strings included terms related to project outcomes (e.g., success, fail, challeng*), contributing factors (e.g., "critical success factor*", model, framework), and specific Agile methodologies (e.g., Agile, scrum, kanban, "extreme programming"). Initial broad searches proved problematic. For example, a basic Google Scholar search for ('project outcome' and agile) yielded 88,900 results, while a more complex string returned 287,000 results. Similarly, a search in the Scopus database on titles, abstracts, and keywords returned over 25,000 results. Similar results were experienced across the other search systems. When looking at the results, it was evident that the balance between the degree of exhaustiveness (sensitivity) and the precision (specificity) of the search was not well balanced [18]. In other words, while the search returned plenty of literature studies (high sensitivity), it produced a lot of irrelevant studies (low specificity). As a result, given the search terms required, which included the use of synonyms as listed above, and after examining the results against search sensitivity and specificity balance, it made sense to implement the search string at the title-only level. Therefore, in this situation, the title-only level search, as mentioned by Bramer et al. [19], Kraus et al. [20], and Iriarte and Bayona [21], is not only valid but appropriate. Appendix A listing the search strings per search system with record count is accessible from the following link: https://doi.org/10.6084/m9.figshare.30819413.

### 2.2  Source Selection and Screening

The search was conducted across five recognized research databases in the information technology field: Scopus, Web of Science, EBSCOhost, ACM Digital Library, and ScienceDirect. Google Scholar was also used for snowballing to identify additional sources from the reference lists of initial papers.

2.2.1 *Selection Criteria.* The selection of studies was guided by the following criteria:

Inclusion Criteria:

1. Research focused on Agile software development methodologies such as XP, Scrum, Crystal Method, DSDM, FDD, Lean Development, Kanban, and Adaptive Software Development.
2. Studies that identify and describe factors contributing to Agile project success, failure, or challenges.
3. Research presenting a framework, model, or method to evaluate Agile projects.
4. Peer-reviewed primary studies containing empirical data, including journal articles, conference papers, and book chapters.
5. No publication date restrictions were applied.

Exclusion Criteria:

1. Research on non-Agile methodologies.
2. Studies focusing on single techniques within Agile, such as pair programming or code refactoring.
3. Publications not available in full text or written in a language other than English.
4. Duplicate papers from different databases or duplicate reports of the same study.

2.2.2 *Screening Results.* As depicted in the Prisma Flowchart (Fig. 1), the process narrowed a large pool of articles to a final selection. The initial search across the selected systems identified 7,214 articles. After 1,213 duplicates were removed, 6,001 articles were screened. A title and abstract screening left 447 eligible articles. Following a full-text review that excluded 394 articles, 53 papers were selected for data extraction and synthesis. Table 1 details the database search results. The search was concluded in August 2024.

**Table 1: Database Search Results**

| Search System | Search Results | With Duplicates removed | After the Title and Abstract Screening | After Full text review |
|---|---|---|---|---|
| Scopus | 4478 | 4136 | 304 | 34 |
| Web of Science | 1324 | 862 | 47 | 9 |
| EBSCOhost | 490 | 265 | 20 | 2 |
| ACM Digital Library | 452 | 437 | 31 | 1 |
| ScienceDirect | 430 | 261 | 14 | 2 |
| Google Scholar (snowballing) | 40 | 40 | 31 | 5 |
| Total | 7214 | 6001 | 447 | 53 |

2.2.3 *Inter-rater Reliability.* To ensure consistency between the two reviewers in this SLR, inter-rater reliability was measured using the Cohen Kappa statistic, where a value of 1 indicates perfect agreement and 0 indicates no agreement [23]-[25]. The process involved several checks: A pilot screening of eight articles resulted in a Cohen Kappa score of 0.54, indicating moderate agreement. For a title and abstract screening of 50 articles, the score was 0.26 (fair agreement). This prompted the reviewers to refine their shared understanding of the selection criteria. The primary researcher then completed the remaining title and abstract screening with this updated knowledge. For the full-text screening, a test on 24 articles yielded a score of 0.81, showing a high level of agreement and confirming the reliability of the screening process.

### 2.3 Quality Assessment

The 53 selected studies underwent a quality assessment (QA) based on a form adapted from the Critical Appraisal Skills Programme and a study by Dybå and Dingsøyr [26]. The QA criteria used are available in Appendix B (https://doi.org/10.6084/m9.figshare.30819413).

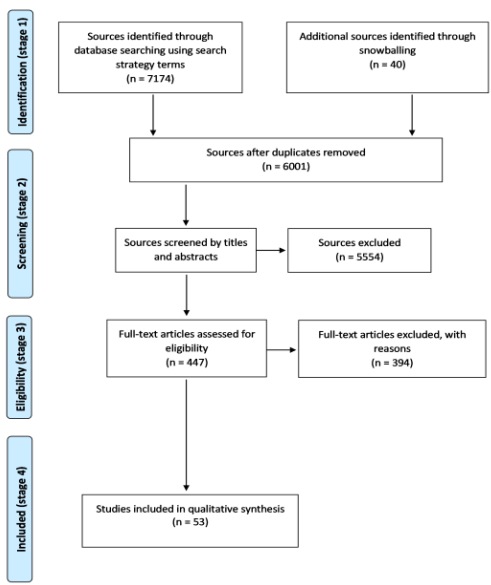

**Figure 1: The Prisma Flowchart stages [22].**

The QA form evaluated studies based on four sections: Reporting, Rigor, Credibility, and Relevance. A scoring system was used for all 11 questions (Yes = 1, Partially = 0.5, No = 0), resulting in a total quality score out of 11. The first three questions assessed basic reporting quality, such as having a clear statement of aims and research design. Studies were excluded if they did not meet all three of these criteria. The remaining eight questions evaluated the quality of the research.

Rigor was assessed by looking at the appropriateness of the research design, recruitment strategy, data collection, and data analysis. Credibility was evaluated by considering ethical issues, the researcher-participant relationship, and the clarity of findings. Finally, relevance was determined by the study's value to research or practice.

### 2.4 Data Extraction

Data was extracted from the 53 finalized studies using a predefined data extraction form within the Covidence SLR tool. To ensure no crucial data was overlooked, the selected papers' findings and conclusions were extracted verbatim, a method that helps in adequately answering the research questions [26].

### 2.5 Synthesis of Findings

The study utilized thematic synthesis with content analysis to synthesize extracted data from the papers included. This approach is effective for integrating qualitative and quantitative data, allowing for both inductive and deductive examination of the text [27], [28]. Thematic synthesis was used to interpret the findings and develop new insights, leading to a proposed theoretical framework [29]. The process consisted of two phases, namely, content analysis of each study, followed by thematic synthesis across all studies. The analysis was performed with human coding in the NVivo software package. The thematic syntheses iterative process is summarized in Appendix C, accessible from the following link: https://doi.org/10.6084/m9.figshare.30819413.

## 3 Results

This section presents the distribution of the selected literature, the taxonomy of synthesized themes and posited critical success factors, and the outcomes of the quality evaluation.

### 3.1 Search Results

A total of 53 articles were selected for this study, with publication dates ranging from 2008 to 2024. As Agile software development being a more recent addition to project management approaches [52], [53], most publications are recent, with most included articles published within the last eight years (2017-2024). No papers from 2010 or 2012 were included in the final selection. The distribution shows that Scopus provided the highest percentage of papers (64%), followed by WoS (17%) and Google Scholar (9%). ScienceDirect and EBSCOhost each contributed 4%, while the ACM Digital Library provided 2%. Regarding research methodology, 55% of the studies were Quantitative, 30% were Qualitative, and 15% used Mixed Methods. Journal articles were the most common publication type, accounting for 77% of the papers, followed by conference papers (19%) and lecture notes (4%).

### 3.2 Quality Evaluation of Articles

The quality of the 53 selected papers was assessed using eleven questions. Appendix D provides the quality evaluation scores assigned to each paper; accessible from the following link: https://doi.org/10.6084/m9.figshare.30819413. The scores ranged from a low of 5 (for one paper) to a high of 10.5 (for one paper), out of a maximum possible score of 11. The average quality score was 7.8, with a median of 7.5 and a standard deviation of 1.16. This average suggests that the papers generally meet most of the quality criteria. No papers were excluded during this stage, as they all met the minimum criteria of the three initial screening questions. It was noted that most papers did not address ethics and reflexivity concerns, which prevented the average score from being higher.

### 3.3 Taxonomy of the Synthesized Findings

Following the thematic synthesis, the extracted data from the included papers were analyzed to create factors from codes. These factors were then organized into five final themes, a process informed by the framework of Chow and Cao [7]. This result was achieved after eight iterations of synthesizing raw codes into the final 21 success factors. The synthesized themes and success factors at the concept level (Appendix E) and SLR source to paper mapping (Appendix F) are accessible from the following link: https://doi.org/10.6084/m9.figshare.30819413. An Agile project success factor taxonomy diagram is displayed in Fig. 2.

### 3.4 Agile Project Success Dimensions

Regarding the dependent variables, which represent the assessment of a project's overall success, Chow and Cao [7] and Stankovic et al. [8] suggest quality (i.e. delivering a good working product), scope (meeting all requirements and objectives), time (delivering on time), and cost (delivering within estimated cost and effort). In addition, Thomas and Fernández [30] and Salman et al. [31] include customer satisfaction (providing a product or project outcome that satisfies the customer), Pereira et al. [32] and Ika and Pinto [33] suggest business goals (the impact the project had on the organization responsible for the delivery of the product). These six comprehensive criteria allow software project success to be evaluated based on three criteria categories [30], **project management** (Time, Cost, Scope), **technical** (Quality, Customer satisfaction), and **business** (Business goals). Thomas and Fernández [30] highlighted that this list of six is slightly above the average amount that companies consider. The rationale for including these criteria is that they have been cited consistently by numerous studies, adding rigor and validity to the measurement and evaluation of Agile project success.

### 3.5 Agile Project Success Theoretical Framework (APSTF)

Depicted in Fig. 3, this SLR resulted in the development of the Agile Project Success Theoretical Framework (APSTF). This framework presents the factors that contribute to Agile project success and the proposed direction of their relationships. It can be used with statistical analysis to identify which factors have a significant linear relationship with project success. In the framework, Agile Project Success is the dependent variable, which is a composite of the dimensions of Quality (Q), Scope (S), Time (T), Cost (C), Customer Satisfaction (K), and Business Goals (B). Success depends on the independent variables, which are the 21 factors grouped into five themes: Technical, Process, Organizational, People, and Project Factors.

### 3.6 Discussion

This section discusses the key factors contributing to Agile software development project success, answering the central research question. The findings identified twenty-one hypothesized critical success factors, which are grouped into five core themes deductively derived from Chow and Cao [7]: 1) Technical Factors, 2) Organizational Factors, 3) People Factors, 4) Project Factors, and 5) Process Factors. Each theme has three to five subthemes (categories) that translate to the success factors. In turn, the factors contain attributes as part of their composition. The discussion will define five themes and then discuss the success factors relative to each theme.

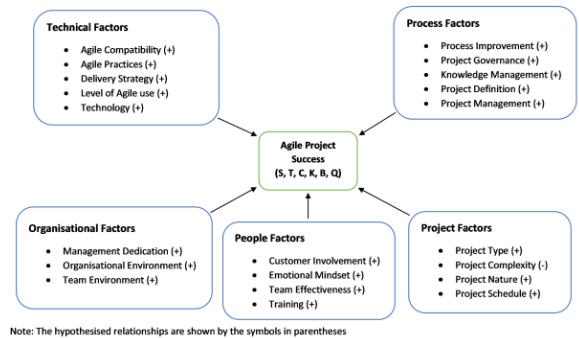

**Figure 3: Agile project success theoretical framework (APSTF).**

**3.6.1 Technical Factors**. This theme includes five factors derived from the SLR: Agile Compatibility, Agile Practices, Delivery Strategy, Level of Agile Use, and Technology.

*Agile Compatibility.* Refers to how consistent a project is with Agile methods and whether it is the most suitable approach [34]. Seven of the 53 papers suggest compatibility is essential for project success (S36, S8, S7, S15, S31, S33, and S38). Concerns were raised about using Agile for incompatible projects, such as IT infrastructure projects with long timescales, with some arguing that Agile "does not apply to every type of project" (S36). Using Agile in an inappropriate setting, like a hierarchical public sector organization, can lead to failure (S15). Overall, Agile Compatibility is seen as a positive influence on project success.

*Agile Practices.* Encompasses a set of practices that help teams navigate the complexities of software development [7], [35]. Sixteen papers were included for this factor (S9, S3, S6, S13, S25, S27, S28, S12, S16, S17, S26, S29, S32, S43, S46, and S53). Examples include daily stand-ups, iterative planning, retrospectives, and pair programming. Studies have identified Agile practices and techniques as critical to project success (S9, S27, S28). It is advised that Agile Practices be included in any evaluation of success factors, as they are seen to influence project outcomes positively.

*Delivery Strategy.* Focuses on how project deliverables are released to the client [35], emphasising the ability to deliver regularly and with the most critical features first [7], [49]. Seven papers were synthesized for this factor (S9, S36, S31, S3, S27, S26, S20). Sithambaram and Nasir's (S36) study with 42 industry practitioners identified improper deliverables as a challenge, proposing that "units of deliverables need to be small enough to control and large enough to create value". Delivery Strategy is viewed as positively influencing project success.

***Level of Agile Use.*** This is the degree of agility in a project achieved by applying the Agile mindset, which goes beyond simply implementing practices [36]. It involves valuing principles like customer collaboration and responding to change [37].

Seven papers suggested that a level of Agile use contributes to Agile outcomes (S8, S26, S14, S35, S47, S48, and S52). Serrador and Pinto's (S35) global quantitative study of 1,002 projects found that the level of Agile use had a statistically significant impact on project success. The study by Moloto et al. (S26) also found a statistically significant relationship between Agile methodology use and project success. This factor is seen as positively influencing the success of Agile projects.

***Technology.*** Refers to resources like Jira, Trello, or video conferencing tools that support the Agile team but are not formal Agile practices themselves [38]. Six papers indicated that Technology plays a role in project success (S2, S28, S16, S14, S21, S43). The inappropriate use of technology, such as obsolete operating systems, was identified as a challenge (S21). Using new, unfamiliar technologies can also create uncertainty and negatively affect project success (S2). Therefore, teams must learn and adapt to employ appropriate technologies (S2, S28). Technology is considered a positive influence on project success.

**3.6.2 Organizational Factors.** This theme consists of three factors derived from the SLR: Management Dedication, Organizational Environment, and Team Environment.

***Management Dedication.*** This factor was created to refer to management's commitment to and support for an Agile project's success. It combines commitment (willingness to implement Agile) and support (providing resources, budgeting, and decision-making) [38]. Fourteen papers contributed to this factor (S36, S13, S15, S3, S27, S12, S16, S20, S14, S21, S18, S43, S47, and S51). Studies show that a lack of management commitment (S15, S21, S36) and executive support (S12, S16, S21) can be significant challenges. Management support is considered critical for allocating necessary resources (S27) and executing the changes needed to transition from traditional methods (S18, S36). This factor positively influences project success.

***Organizational Environment.*** This refers to the structure and culture of the organization [7]. Nineteen papers were included in this factor (S2, S36, S15, S3, S27, S16, S20, S14, S21, S18, S1, S5, S10, S30, S41, S42, S13, S43, S49). A traditional, hierarchical structure can be a challenge when teams adopt Agile (S15). A flexible, less rigid organizational culture is ideal for Agile projects to succeed (S18). A culture that is too traditional has been identified as a critical failure factor (S30). The environment also includes performance evaluation, which should shift from individual to team-based to improve collaboration (S10). Organizational environment positively influences project success.

***Team Environment.*** This factor relates to team logistics, such as the collocation of members, team diversity, and team structure [7], [38]-[41]. Ten papers were included in this factor (S9, S15, S20, S21, S1, S41, S13, S43, S46, and S25). Chow and Cao (S9) recognized the team environment as a CSF contributing to project quality. An Agile-friendly team environment positively affects success (S20). Team logistics (S1, S18, S21, S41), such as workspace arrangement (S18, S41) and avoiding over-time (S41), are also key contributors. This factor is seen as positively influencing project success.

**3.6.3 People Factors.** This theme contains four factors: Customer Involvement, Emotional Mindset, Team Effectiveness, and Training.

***Customer Involvement.*** This involves the customer's relationship with the team, their level of communication, and their commitment to the project [7]. Twenty of the 53 papers were included (S2, S9, S36, S3, S28, S16, S29, S20, S21, S18, S1, S5, S30, S13, S23, S43, S45, S34, S47, and S50). Studies show customer involvement is a significant factor in project success, particularly regarding scope (S9). A good customer relationship is vital for getting support and buy-in (S36), while customer commitment is also essential for positive project outcomes (S9, S28, S30). Customer Involvement is seen to positively influence project success.

***Emotional Mindset (EM).*** EM as an Agile project success factor has not been featured in the synthesized papers, albeit Emotional Intelligence (EI) has. The study by Luong et al. [54] focused on EI's role in the outcomes of Agile information systems projects, suggesting that EI plays a vital role in Agile teams and the workplace.

To understand EM in the context of this study, you need to understand the concepts of EI and Mindset. Salovey and Mayer [55] define EI as "The ability to monitor one's own and others' feelings and emotions, to discriminate among them and to use this information to guide one's thinking and actions". On the other hand, Mindset refers to attitudes, beliefs, and thought patterns that influence your behavior, decisions, and response to challenges and situations [56], [57]. Therefore, the two concepts differ based on the definitions of EI and mindset described above. At the same time, EI and Mindset involve an individual's self-awareness, attitudes, beliefs, behaviors, and how they perceive and interact in society, including the workplace. EI focuses on the emotional and social aspects of the individual. In contrast, Mindset focuses on the individual's cognitive and mental elements, such as attitudes, beliefs, and thought patterns. EM within the context of this study is a combination of the EI and Mindset concepts. EM refers to an individual's ability to respond and interpret their own and others' emotions, including having a set of beliefs, attitudes, and behaviors that shape their emotional experiences and feelings [42]. The factor was synthesized from 18 papers (S2, S34, S36, S15, S33, S12, S29, S14, S21, S18, S10, S41, S42, S24, S19, S22, S37, and S51) covering attributes of human behavior (e.g., commitment to work) (S2, S12, S14, S21, S19, S22, S37), emotion (e.g., stress, fear) (S10, S29), and mindset (e.g., motivation, attitude) (S34, S36, S15, S33, S18, S41, S42, S51, S106). Studies have shown a positive relationship between motivation (S34, S18) and project success, and that having the right team attitude (S41) plays an important role in project outcomes. EM is posited to positively influence project success.

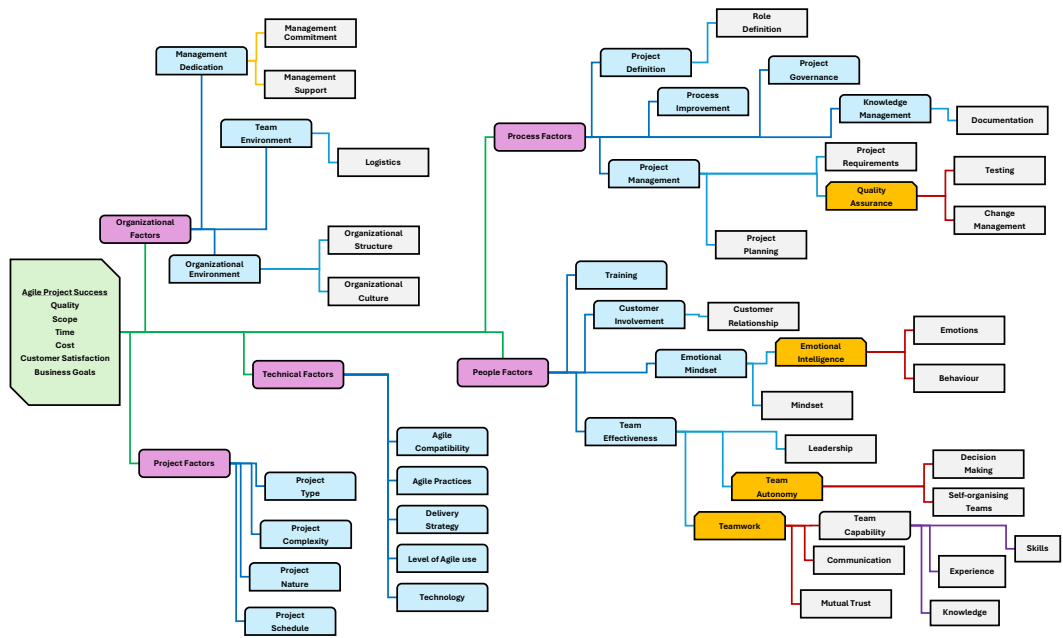

**Diagram Five Colour Scheme**: 1) **Green** – The Agile project success criteria to measure success in terms of Q, S, T, C, K, B, and Q. 2) **Purple** – The five themes of Organizational, Technical, Project, People, and Process Factors under which synthesized factors are placed. 3) **Blue** – The 21 factors hypothesized to contribute to Agile project success. 4) **Orange** – Sub-categories of a factor, which can be seen as an attribute with sub-attributes (grey). 5) **Grey** – Attributes of the factor or its sub-category

**Figure 2: Agile project success taxonomy diagram.**

*Team Effectiveness.* This factor refers to individuals working interdependently and cooperatively toward a shared goal [43]. Synthesized from 37 (S2, S9, S34, S36, S15, S33, S3, S27, S12, S16, S17, S29, S20, S14, S21, S18, S1, S5, S10, S30, S41, S42, S19, S22, S37, S40, S44, S4, S45, S23, S46, S47, S51, S13, S45, S43, S11) of the 53 papers, it is the most cited factor in the SLR. Its attributes include teamwork (S36, S15, S27, S16, S1, S22, S51), team capability and communication (S9, S13, S34, S36, S45, S47, S15, S3, S27, S16, S17, S29, S20, S14, S21, S1, S5, S10, S41, S42, S24, S19, S22, S37, S40, S43, S44, S51), team autonomy (S15, S12, S18, S37, S4, and S47), and leadership (S2, S15, S23, S33, and S27). Research shows that teamwork quality (S22) has a positive effect on team performance and that a lack of teamwork (S16, S36) is a significant impediment. Team capability (competence and expertise) is also a significant explainer of project success (S9, S15, S3, S17, S29, S41, S19, and S40). Team autonomy, or self-organizing teams, positively impacts project success (S15, S42) by enabling decision-making at the team level (S4, S37). Leadership styles like transformational and shared leadership are also aligned with project success (S15, S33). This factor is seen as positively influencing project success.

*Training.* A lack of proper training on Agile practices is posited to affect project success negatively [44]. Six studies were included for this factor (S16, S29, S21, S10, S13, and S38). A training deficit can lead to ignoring Agile values and contribute to project failure (S16). However, attending training is not enough, in other words, it is the correct application of what was learned that matters (S21 and S10). Continuous training is essential for understanding Agile principles (S10). Training is viewed as positively influencing project success.

**3.6.4 Project Factors**. This theme covers four factors: Project Type, Project Complexity, Project Nature, and Project Schedule.

*Project Type.* This concerns the suitability of a project for Agile, such as projects with variable scope and emergent requirements [7], [38]. Two of the 53 synthesized studies were included for this factor (S20 and S41). Kulathunga and Ratiyala's (S20) found that projects with variable scopes positively affect success. It's suggested that there is a need to match the project type with the right management approach, as Agile may not be suitable for all types, like legacy systems (S41). Project Type is seen as a positive influence on success.

*Project Complexity.* Refers to the technical complexity of project tasks, applications, and systems. This is distinct from project size [3], [45], [46]. Five papers were included for this factor (S2, S14, S21, S13, and S47). High technical complexity can cause the implementation process to deteriorate, leading to projects being behind schedule and over budget (S2). Therefore, Project Complexity is seen as negatively influencing project success.

*Project Nature.* This relates to whether a project is "life-critical" (e.g., advanced weapons systems) [47] or not [7]. Four of the 53 synthesized studies were included for this factor (S39, S20, S14, and S13). Stankovic et al. (S39) found that project nature significantly influences whether a project will finish on time and within budget. Kulathunga and Ratiyala (S20) found that limiting projects to non-life-critical ones had a positive effect on success. Project Nature is seen as positively influencing success.

*Project Schedule.* In Agile, this refers to a dynamic, accelerated schedule that aligns with the principle of delivering working software frequently [7]. Four of the 53 synthesized studies were included for this factor (S39, S36, S20, and S13). Stankovic et al. (S39) found that the project schedule was a critical success factor influencing project cost. It is posited that controlling the schedule is essential, while Kulathunga and Ratiyala (S20) accepted the hypothesis that dynamic schedules positively affect Scrum project success. Project Schedule is viewed as a positive influence.

**3.6.5 Process Factors**. This theme consists of five factors: Process Improvement, Project Governance, Knowledge Management, Project Definition, and Project Management.

*Process Improvement*. This involves applying a structured approach to improve project outcomes, boost efficiency, and reduce waste [48]. Three of the 53 synthesized studies were included for this factor (S15, S36, S38). Gregory et al. (S15) identified process improvement as the second most mentioned challenge by practitioners, who noted that Agile demands ongoing transformation to become sustainable. Process Improvement positively influences project success.

*Project Governance*. This is the framework that monitors and controls project variables like policies, regulations, and responsibilities [49]. Four of the 53 synthesized studies were included for this factor (S36, S15, S4, and S43). Practitioners have struggled to see the link between governance and project outcomes, but it is a significant part of projects in regulated environments like banking (S15). Awais et al. (S4) found that the clarity of decision-making regarding governance is positively related to project success. Project Governance is seen as a positive influence.

*Knowledge Management.* This focuses on effectively capturing, organizing, and using knowledge, including documentation [50]. Five studies were included (S15, S12, S21, S19, and S44). Lack of information use in sprints can be caused by the extra time and effort required to use it (S19). While Agile teams may not use much documentation, preserving some records is vital for future reference (S12). Efficient knowledge management has been shown to positively impact project success (S44). This factor is seen as a positive influence.

*Project Definition.* Referring to a well-defined project with upfront details, including cost evaluation and risk analysis [7]. Chow and Cao's [7] study did not find this factor critical. However, a replication study by Stankovic et al. [8] did, particularly regarding project cost. Five of the 53 synthesized studies were included for this factor (S15, S12, S21, S19, and S44). Lack of a defined scope has been found to be a critical failure factor (S1). The clear definition of roles and responsibilities is also essential for smooth project delivery (S36). Project Definition is seen to positively influence success.

*Project Management.* Involves applying techniques, skills, and tools to meet project requirements [51]. Synthesized from 22 papers (S2, S9, S39, S36, S31, S3, S27, S12, S26, S29, S20, S14, S35, S21, S1, S41, S42, S11, S23, S47, S50, S13), its attributes include quality assurance (QA), requirements management, and project planning. Studies have proven the significance of the project management process to success in terms of quality (S9, S12) and cost (S39). Proper project planning is a CSF for delivering a project on time (S14). QA tasks like testing (S12, S21) and change management (S2, S21, S36) are also critical. For instance, bugs found late can interrupt planned iterations (S12). Project Management positively influences project success.

## 4   Conclusions

The research question, "What are the key factors contributing to Agile software development project success?" was answered by conducting an SLR that included 53 papers and used thematic synthesis with content analysis. Based on the findings, 21 factors are hypothesized to contribute significantly to Agile project success. These factors are management dedication, organizational environment, team environment, customer involvement, emotional mindset, team effectiveness, training, knowledge management, process improvement, project definition, project governance, project management, project complexity, project nature, project schedule, project type, Agile compatibility, Agile practices, delivery strategy, level of Agile use, and technology. Furthermore, these 21 factors are grouped into five themes: organizational, people, technical, process, and project. Team effectiveness and project management are the most coded factors in this study. These two critical factors are anticipated to contribute significantly to Agile software development success.

In addition, developing the Agile Project Success Theoretical Framework (APSTF) for further research investigation could replicate this SLR to approve (or debunk) our findings. Furthermore, a quantitative survey could identify the factors that positively correlate with agile project success, practically apply the findings to an Agile project, and evaluate the contributions.

Limitations of this study include a lack of resources and time. First, there is a lack of resources to conduct the SLR. It would have been preferential if the team had more researchers, potentially contributing to additional insights. Second, the lack of time prevents us from longitudinally continuing this SLR, as an example, allowing for more time to add additional research to the data extraction and synthesis even after writing up the findings.

## References

[1] Ahimbisibwe, A., Cavana, R. Y., Daellenbach, U.: A contingency fit model of critical success factors for software development projects: A comparison of agile and traditional plan-based methodologies. Journal of Enterprise Information Management, 28(1), pp. 7-33 (2015), doi: 10.1108/JEIM-08-2013-0060

[2] Binboga, B., Gumussoy, C. A.: Factors affecting agile software project success. IEEE Access, 12, pp. 95613-95633 (2024), doi: 10.1109/ACCESS.2024.3384410

[3] Ahimbisibwe, A., Daellenbach, U., Cavana, R. Y.: Empirical comparison of traditional plan-based and agile methodologies: Critical success factors for outsourced software development projects from vendors' perspective. Journal of Enterprise Information Management, 30(3), pp. 400-453 (2017) doi: 10.1108/JEIM-06-2015-0056

[4] Hanslo, R., Tanner, M.: Machine Learning models to predict Agile Methodology adoption. In: 15th Conference on Computer Science and Information Systems (FedCSIS), IEEE, pp. 697-704 (2020)

[5] Chekmarev, A. V.: Agile with relation to Conflicts Theory. In: Proceedings of the 2019 IEEE International Conference of Quality Management,

Transport and Information Security, Information Technologies IT and QM and IS, IEEE, pp. 23-26 (2019)

[6] Gemino, A., Horner Reich, B., Serrador, P. M.: Agile, traditional, and hybrid approaches to project success: is hybrid a poor second choice?. Project management journal, 52(2), pp. 161-175 (2021)

[7] Chow, T., Cao, D. B.: A survey study of critical success factors in agile software projects. Journal of Systems and Software, 81(6), pp. 961-971 (2008), doi: 10.1016/j.jss.2007.08.020

[8] Stankovic, D., Nikolic, V., Djordjevic, M., Cao, D. B.: A survey study of critical success factors in agile software projects in former Yugoslavia IT companies. Journal of Systems and Software, 86(6), pp. 1663-1678 (2013), doi: 10.1016/j.jss.2013.02.027

[9] Rockart, J. F.: Chief executives define their own data needs. Harvard business review, 57(2), pp. 81-93 (1979)

[10] Fink, A.: Conducting research literature reviews: From the internet to paper. Sage publications (2019)

[11] Keele, S.: Guidelines for performing systematic literature reviews in software engineering. Tech. Rep. ver. 2.3 ebse (2007)

[12] Rowe, F.: What literature review is not: diversity, boundaries and recommendations. European Journal of Information Systems, 23(3), pp. 241-255 (2014)

[13] Vom Brocke, J., Simons, A., Riemer, K., Niehaves, B., Plattfaut, R., Cleven, A.: Standing on the shoulders of giants: Challenges and recommendations of literature search in information systems research. Communications of the Association for Information Systems, 37(1), 9 (2015)

[14] Kitchenham, B.: Procedures for performing systematic reviews. Keele, UK, Keele University, 33, pp. 1-26 (2004)

[15] Paré, G., Trudel, M. C., Jaana, M., Kitsiou, S.: Synthesizing information systems knowledge: A typology of literature reviews. Information and Management, 52(2), pp. 183-199 (2015)

[16] Schryen, G.: Writing qualitative IS literature reviews – Guidelines for synthesis, interpretation and guidance of research. Communications of the Association for Information Systems, 37(12), pp. 286-325 (2015)

[17] Oosterwyk, G., Brown, I., Geeling, S.: A Synthesis of Literature Review Guidelines from Information Systems Journals. In: Proceedings of 4th International Conference on the Internet, Cyber Security and Information Systems (ICICIS), 12, pp. 250-260 (2019)

[18] Xiao, Y., Watson, M.: Guidance on conducting a systematic literature review. Journal of planning education and research, 39(1), pp. 93-112 (2019)

[19] Bramer, W. M., Giustini, D., Kleijnen, J., Franco, O. H.: Searching Embase and MEDLINE by using only major descriptors or title and abstract fields: a prospective exploratory study. Systematic Reviews, 7, pp. 1-8 (2018)

[20] Kraus, S., Breier, M., Dasí-Rodríguez, S.: The art of crafting a systematic literature review in entrepreneurship research. International Entrepreneurship and Management Journal, 16, pp. 1023-1042 (2020)

[21] Iriarte, C., Bayona, S.: IT projects success factors: a literature review. International Journal of Information Systems and Project Management, 8(2), pp. 49-78 (2020), doi: 10.12821/ijispm080203

[22] Liberati, A., Altman, D. G., Tetzlaff, J., Mulrow, C., Gøtzsche, P. C., Ioannidis, J. P., Clarke, M., Devereaux, P. J., Kleijnen, J., Moher, D.: The PRISMA statement for reporting systematic reviews and meta-analyses of studies that evaluate health care interventions: explanation and elaboration. Annals of internal medicine, 151(4), pp. 65-94 (2009), doi: 10.7326/0003-4819-151-4-200908180-00136

[23] Cicchetti, D. V.: Guidelines, criteria, and rules of thumb for evaluating normed and standardized assessment instruments in psychology. Psychological assessment, 6(4), 284 (1994).

[24] Fleiss, J. L.: Measuring nominal scale agreement among many raters. Psychological bulletin, 76(5), 378 (1971)

[25] McHugh, M. L.: Interrater reliability: the kappa statistic. Biochemia medica, 22(3), pp. 276-282 (2012)

[26] Dybå, T., Dingsøyr, T.: Empirical studies of agile software development: A systematic review. Information and Software Technology, 50(9), pp. 833-859 (2008), doi: 10.1016/j.infsof.2008.01.006

[27] Neuendorf, K. A.: Content analysis and thematic analysis. In: Advanced research methods for applied psychology, pp. 211-223 (2018)

[28] Vaismoradi, M., Turunen, H., Bondas, T.: Content analysis and thematic analysis: Implications for conducting a qualitative descriptive study. Nursing and health sciences, 15(3), pp. 398-405 (2013)

[29] Bearman, M., Dawson, P.: Qualitative synthesis and systematic review in health professions education. Medical education, 47(3), pp. 252-260 (2013)

[30] Thomas, G., Fernández, W.: Success in IT projects: A matter of definition?. International Journal of Project Management, 26(7), pp. 733-742 (2008)

[31] Salman, A., Jaafar, M., Malik, S., Mohammad, D., Muhammad, S. A.: An Empirical Investigation of the Impact of the Communication and Employee Motivation on the Project Success Using Agile Framework and Its Effect on the Software Development Business. Business Perspectives and Research, 9(1), pp. 46-61 (2021), doi: 10.1177/2278533720902915

[32] Pereira, J., Varajão, J., Takagi, N.: Evaluation of information systems project success‐Insights from practitioners. Information Systems Management, 39(2), pp. 138-155 (2022)

[33] Ika, L. A., Pinto, J. K.: The 're-meaning' of project success: Updating and recalibrating for a modern project management. International Journal of Project Management, 40(7), pp. 835-848 (2022)

[34] Moore, G. C., Benbasat, I.: Development of an instrument to measure the perceptions of adopting an information technology innovation. Information Systems Research, 2(3), pp. 192-222 (1991)

[35] Muhammad, A., Siddique, A., Naveed, Q. N., Saleem, U., Hasan, M. A., Shahzad, B.: Investigating crucial factors of agile software development through composite approach. Intelligent Automation and Soft Computing, 27(1), pp. 15-34 (2021), doi: 10.32604/iasc.2021.014427

[36] van Kelle, E.: Social Factors of Agile Development Success. M.S. thesis, Tilburg University., Tilburg, Netherlands (2014)

[37] Beck, K., Beedle, M., Van Bennekum, A., Cockburn, A., Cunningham, W., Fowler, M., Grenning, J., Highsmith, J., Hunt, A., Jeffries, R.: The agile manifesto (2001)

[38] Cao, D. B.: An empirical investigation of critical success factors in agile software development projects. Ph.D. dissertation, Capella University (2006)

[39] Tanner, M., and von Willingh, U.: Factors leading to the success and failure of agile projects implemented in traditionally waterfall environments. Human capital without borders: Knowledge and learning for the quality of life. Portoroz, Slovenia: Make Learn, pp. 693-701 (2014)

[40] Darnell, R. C.: Implicit and explicit measures of coordination effectiveness as predictors of agile software development project success: A regression approach. Ph.D. dissertation, Capella University (2015)

[41] Meurs, T. T., Gevers, J., Le Blanc, P.: Determining the success factors of Scrum projects. M.S. thesis. University of Technology, Netherlands (2015)

[42] Fredrickson, B. L., Losada, M. F.: Positive affect and the complex dynamics of human flourishing. American psychologist, 60(7), 678 (2005)

[43] Salas, E., Rico, R., Passmore, J.: The psychology of teamwork and collaborative processes. The Wiley Blackwell handbook of the psychology of team working and collabo-rative processes, pp. 1-11 (2017)

[44] Nisyak, A. K., Rizkiyah, K., Raharjo, T.: Human related challenges in agile software development of government outsourcing project. In 7th International Conference on Electrical Engineering, Computer Sciences and Informatics (EECSI), IEEE, pp. 222-229 (2020)

[45] Garousi, V., Tarhan, A., Pfahl, D., Coşkunçay, A., Demirörs, O.: Correlation of critical success factors with success of software projects: an empirical investigation. Software Quality Journal, 27(1), pp. 429-493 (2019)

[46] Kunda, D., Mulenga, M., Sinyinda, M., Chama, V.: Challenges of Agile development and implementation in a developing country: A Zambia case study. Journal of Comput-er Science, 14(5), pp. 585-600 (2018), doi: 10.3844/jcssp.2018.585.600

[47] Brown, G. A.: An examination of critical success factors of an agile project. Ph.D. dissertation, Capella University (2015)

[48] Farlik, J. T.: Project success in agile development software projects. Ph.D. dissertation, Capella University (2016)

[49] Sithambaram, J., Nasir, M. H. N. B. M., Ahmad, R.: Issues and challenges impacting the successful management of agile-hybrid projects: A grounded theory approach. International Journal of Project Management (2021), doi: 10.1016/j.ijproman.2021.03.002

[50] Despres, C., Chauvel, D.: Knowledge management (s). Journal of knowledge Management (1999)

[51] Schwalbe, K.: Introduction to project management. Course Technology, Cengage Learning, Boston (2009)

[52] Nguyen, D. S.: Success factors that influence agile software development project success. American Scientific Research Journal for Engineering, Technology, and Sciences (ASRJETS), 17(1), pp. 171-222 (2016)

[53] Tam, C., Moura, E. J. D. C., Oliveira, T., Varajão, J.: The factors influencing the success of on-going agile software development projects. International Journal of Project Management, 38(3), pp. 165-176 (2020), doi: 10.1016/j.ijproman.2020.02.001

[54] Luong, T. T., Sivarajah, U., Weerakkody, V.: Do Agile Managed Information Systems Projects Fail Due to a Lack of Emotional Intelligence? Information Systems Frontiers, 23(2), pp. 415-433 (2021), doi: 10.1007/s10796-019-09962-6

[55] Salovey, P., Mayer, J. D.: Emotional intelligence. Imagination, cognition and personality, 9(3), pp. 185-211 (1990)

[56] Burnette, J. L., Finkel, E. J.: Buffering against weight gain following dieting setbacks: An implicit theory intervention. Journal of Experimental Social Psychology, 48(3), pp. 721-725 (2012)

[57] Dweck, C. S.: Mindset: The new psychology of success. Random house (2006)